\documentclass{svmult}                                        
                                                              
\usepackage{makeidx}     % allows index generation            
\usepackage{graphicx}    % standard LaTeX graphics tool       
\usepackage{multicol}    % used for the two-column index      
\usepackage{amsfonts,amsmath,amssymb,latexsym}                
\begin{document}

\title*{Tree decomposition  and postoptimality analysis in discrete
optimization\thanks{Research supported by FWF (Austrian Science
Funds) under the project P20900-N13.}
}

\titlerunning{Tree decomposition and postoptimality analysis}        % if too long for running head

\author{Oleg Shcherbina}

%\inst{1}
%\and
%Name of Author\inst{2}}
\institute{Faculty of Mathematics,\\ University of Vienna\\
%\at
 Nordbergstrasse 15,  A-1090 Vienna,\\ Austria \\
\texttt{oleg.shcherbina@univie.ac.at}}
%\and Name and Address of your Institute \texttt{name@email.address}
\maketitle


\begin{abstract}
Many real discrete optimization problems (DOPs)
are $NP$-hard and contain a huge number of variables and/or constraints that make the
models intractable for currently available  solvers.  Large DOPs can be
solved due to their special structure using decomposition approaches.
 An important example of decomposition approaches is tree decomposition with
  local decomposition algorithms using the special block matrix structure of constraints which
 can exploit sparsity in the interaction graph of a discrete
optimization problem. In this paper,  discrete optimization problems  with a
tree structural graph are solved by local decomposition
algorithms. Local decomposition algorithms generate a family of related
DO problems which have the same structure but differ in the right-hand sides.
Due to this fact,  postoptimality techniques in DO are applied.
%\keywords{discrete optimization  \and decomposition \and  tree decomposition
%\and postoptimal analysis}
%\subclass{90C10 \and 90C35 \and 90C39}
\end{abstract}

\section{Introduction} Discrete optimization (DO) problems  arise
in various application areas such as planning,
economical allocation, logistics, scheduling, computer aided design,
and robotics.
Application areas od discrete optimization models of OR also include 
supply chain design and management, network optimization,
telecommunications,
VLSI routing, manufacturing, transportation, scheduling, and
finance.  The tremendous
attention that DO particularly has received in the literature gives
some indication of its importance in many research areas.
Unfortunately, most of the interesting problems are in the
complexity class $NP$-hard and may require searching a tree of
exponential size (if $P \neq NP$) in the worst case. Many real DO
problems contain a huge number of variables and/or constraints that
make the models intractable for currently available DO solvers.

One of the promising approaches to cope with NP-hardness in solving
DO problems is the construction of decomposition methods \cite{RalG05}.
Decomposition techniques usually determine
subproblems the solutions of which can be combined to create a solution of
the initial DO problem. Usually, DO problems from  applications have
a special structure, and the matrices of constraints for large-scale
problems have a lot of zero elements (sparse matrices). The nonzero
elements of the matrix often fall into a limited number of blocks.
The block form of many DO problems is usually caused by the weak
connectedness of subsystems of real-world systems.

The search for graph structures appropriate for the application of
dynamic programming caused a series of papers dedicated to tree
decomposition research (\cite{RobSey},  \cite{Arn91}, \cite{Blair},
\cite{Kloks}, \cite{Dechter2001}, \cite{Bodl93}, \cite{bodl},
 \cite{Hegg}). Tree decomposition methods aim to merge
variables such that the meta-graph is a tree of meta-nodes. Tree decomposition and the related
notion of a \textbf{treewidth} ({\sc Robertson, Seymour} \cite{RobSey}) play
a very important role in algorithms, for many $NP$-complete problems
on graphs that are otherwise intractable become polynomial time
solvable when these graphs have a tree decomposition with restricted
maximal size of cliques (or have a bounded treewidth \cite{Bodl93},
\cite{bodl}).

Most of the works based on tree decomposition approach only present theoretical results
\cite{Jegou_dyn}, see the recent survey by {\sc Hicks et al.}
\cite{Hicks}. But only few papers on applications of this powerful tool in the
area of DO exist \cite{Koster99}, \cite{Hicks}.

The algorithmic importance of the tree decomposition was caused by results of
{\sc Courcelle}
\cite{cour} and {\sc Arnborg} et al. \cite{Arn87} which showed that
several $NP$-hard problems posed in monadic second-order logic can be solved in polynomial
time using dynamic programming techniques on input graphs with bounded treewidth

Tree decomposition based algorithms demonstrated their efficiency
on solving frequency assignment problem {\sc{Koster et al.}}
\cite{Koster99},  ring-routing problems \cite{cook94}, and traveling salesman
problem \cite{cook03}.

Efficiency of tree decomposition based  algorithms crucially depends on interaction
graph structure of the DO problem, so that it has a time complexity  $O(n \cdot 2^{tw+1})$,
where $tw$ is the treewidth of the graph.
 If the interaction graph is rather sparse
or, in other words, it has a relatively small treewidth, then
complexity of the tree decomposition algorithm is reasonable.

 Necessity of reduction of enumeration while solving problems corresponding to meta-nodes
 of the tree decomposition causes expediency and an urgency of development of tools that could
 help to cope with this difficulty.

In this paper,  discrete optimization problems  with a
tree structural graph are solved by local decomposition
algorithms that belong to dynamic programming paradigm. Local decomposition algorithm generates
a family of related
DO problems which have the same structure but differ in the right-hand sides.
Due to this fact, postoptimality techniques in DO are applied.

\section{ Discrete optimization problems with constraints and their
graph representations}

Consider a DOP with constraints:
\begin{eqnarray}\label{goal1}
\max_X f(X) = \max_X \sum_{k \in K} f_k (X^k),
\end{eqnarray}
subject to\\
\begin{equation}\label{cons1} A_{iS_i}X_{S_i} \leq
b_i,~  i \in M = \{ 1, 2, \ldots, m \},
\end{equation}
\begin{equation}
\label{cons2} x_{j}=0,1,~ j \in N = \{ 1,\ldots, n
\}, \end{equation}
where\\
$X = \left\{ x_1,\dots,x_n \right\}$ is a
set of discrete variables, functions $f_i(X^i)$ are called
components of the objective function and can be defined in tabular
form, $X^k \subset X,~~k \in K=\left\{1,2,\ldots,t \right\},$ $t$ is the 
number of components of objective function, $K$ is a set of indices
of components;
\begin{eqnarray}\label{sub}
S_{i}\subseteq \{ 1,2, \ldots, n\},~ i \in M.
\end{eqnarray}

We shall consider further a linear objective function (\ref{goal2}):
\begin{eqnarray}\label{goal2} F(x_{1}, \ldots, x_{n})= F(X)= C_N
X_N=\sum_{j=1}^n c_jx_j \rightarrow \max
\end{eqnarray}

\begin{definition}\cite{BerBri}. Variables $x \in X$ and $y \in Y$
interact in  DOP with constraints if they both appear either in the
same component of objective function, or in the same constraint (in
other words, if variables are both either in a set $X^k$, or in
a set $X_{S_i}$). \end{definition}

Graph representation of a DOP structure may be done with various
detailization. Structural graph of a DOP defines which variables are
in which constraints.
%\begin{definition}\cite{BerBri}. Variables $x \in X$ and $y \in Y$
%\emph{interact} in unconstrained DOP, if there is such component of
%objective function $f_{k}(X^{k})$, that both variables $x$ and $y$
%are in  $X^{k}$ (in other words, variables $x$ and $y$ are adjacent
%on the component $X^{k}$). %\end{definition}

An \textbf{interaction graph} \cite{BerBri}
 represents a
structure of the DOP in a natural way.
\begin{definition}
\cite{BerBri}. The \textbf{interaction graph} of a DOP is an
undirected graph $G=(X, E)$, such that
\begin{itemize} \item[1.]
Vertices $X$ of $G$ correspond to variables of DOP;
\item[2.] Two
vertices of $G$ are adjacent iff corresponding variables interact.
\end{itemize}
\end{definition}
Further, we shall use the notion of
vertices that correspond one-to-one to variables.
\begin{definition}
The set of variables interacting with a variable $x \in X$,
 is denoted by $Nb(x)$ and called \textbf{neighborhood} of the
 variable $x$.
For corresponding vertices a neighborhood of a vertex $v$
 is a set of vertices of interaction graph that are linked by
 edges with $v$. Denote the latter neighborhood as $Nb_G(v)$.
\end{definition}
Let $S$ be a vertex set of the
graph. Introduce the following
notions:
\begin{enumerate}
 \item Neighborhood of a set $S
\subseteq V$, $Nb(S) = \bigcup_{v \in S}Nb(v)-S $.
\item Closed neighborhood of a set $S \subseteq V$, $Nb[S]= Nb(S)
\cup S$.
\item If $S=\{j_1,\ldots,j_q\}$ then
$X_S=\{x_{j_1},\ldots,x_{j_q}\}$.
%\item Row-neighborhood of an index
%set $J \subseteq V$:\\ $ U(J) = \{ i | a_{ij} \neq 0, \, j \in J
%\}$, where $A=\|a_{ik}\|$ is a constraint matrix in DOP
%(\ref{goal2}), (\ref{cons1}), (\ref{cons2});
%\item Column-neighborhood of an index set $I \subseteq M = \{ 1, \ldots , m
%\}$:\\ $S(I) = \left\{ k | a_{lk} \neq 0, \, l \in I \right\}$.
%\item A vertex $v$ is {\it simplicial} if $Nb(v)$ is a clique (i.e.
%a set of parwise adjacent vertices).
\end{enumerate}
%\begin{definition} A partition of a set $S$ is a
%decomposition of $S$ into pairwise disjoint nonempty subsets whose
%union is all of $S$.
%\end{definition}
%\begin{definition}\cite{Harary}.
%Given graph $G=(V,E)$, let $V$ be partitioned into
%subsets $S_1,S_2,\ldots, S_k$. The \textbf{condensation}  of $G$
%with respect to this partition is the graph (called condensed
%meta-graph), every $S_p$ is a meta-node and an edge of the condensation 
%graph is defined by the rule: there is an edge from meta-node $S_p$ to
%meta-node $S_q$ in the new graph  iff in $G$ there is at least one
%edge from a meta-node $S_p$ to one of $S_q$.
%\end{definition}
\begin{example}
\[
\begin{aligned}
&2x_1+3x_2+~x_3+5x_4+4x_5+6x_6+~x_7&\rightarrow \max\\
&3x_1+4x_2+~x_3~~~~~~~~~~~~~~~~~~~~~~~~~~&\leq 6,~&(C_1) \\
&~~~~~~~~2x_2+3x_3+3x_4~~~~~~~~~~~~~~~~~~&\leq 5,~&(C_2)\\
&~~~~~~~~2x_2~~~~~~~~~~~~~~~~+3x_5~~~~~~~&\leq 4,~&(C_3)\\
&~~~~~~~~~~~~~~~~2x_3~~~~~~~~~~~~~~~~+3x_6+2x_7~&\leq 5,~&(C_4)\\
&~x_j=0,1,~j=1,\ldots, 7.~~~~~~~~~~~~~~~~~~~~~~~~~~~~~~~~~
\end{aligned}
\]
\end{example}
%\begin{figure}[htbp]
%\centering
%\includegraphics[height=5cm]{twim_elimin_ris1.eps} \caption{Dual
%graph for example 1.} \label{fig:4}
%\end{figure}

\begin{figure}[htbp]
\centering
\includegraphics [height=7cm]{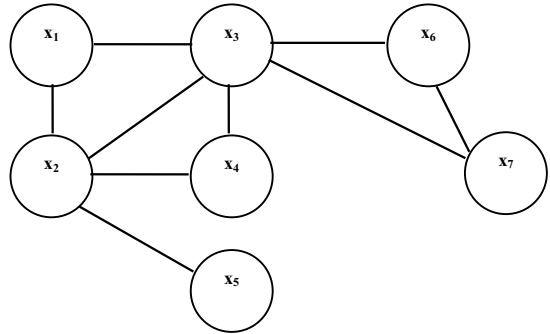}
\caption{Interaction graph for
example.} \label{InterGraph}
\end{figure}
We need following notions.
A \textbf{clique} is a set of vertices that
induce a complete subgraph of $G$, and a \textbf{maximal clique} is a clique
which is not a subset of any other clique.
A clique $C$ in graph $G$ is \textbf{maximal}, if
$C$ is not a subset of any other clique in $G$.
\textbf{A spanning tree} of a
graph is a tree that includes every vertex in  the graph.
 The \textbf{clique graph} of G, is the intersection graph
of the family of maximal cliques  of $G$.

\subsection{Tree decomposition in discrete optimization. Chordal
graphs. Triangulation} \label{Tree}

Dynamic programming is a very powerful algorithmic framework in
which an optimization problem is solved by identifying a set of
subproblems and solving them one-by-one, smallest first, storing
solutions and using the stored solutions to small problems to find
solutions of larger ones (recursively defining value of optimal
solution), until the whole set of them are solved. Dynamic
programming computes recurrences efficiently by storing partial
results in tables.

A tree is a recursive data structure because each child of a node in
the tree has a tree of its descendants. Due to this fact, many of
the important algorithms to access and manipulate trees are easily
expressed using recursion and particularly dynamic programming which
computes recurrences efficiently by storing partial results in
tables and hence can be effectively used for solving optimization
problems on trees.
Dynamic programming works best
on objects which are linearly ordered, e.g. the left-to-right order
of leaves in a tree. Dynamic programming starts at the leaves of the tree and
proceeds from smaller to larger subproblems (corresponding to
subtrees) that is to say, bottom-up in the rooted tree.

One reason why many optimization problems that are hard on general
graphs are easy on trees is that trees do not contain cycles.

Above mentioned  facts and an observation that many optimization
problems which are hard to solve on general graphs are easy on trees
makes the detection tree structures in a graph a very promising
task. A powerful tool of algorithmic graph theory like 
\textbf{tree decomposition} \cite{RobSey} can help detect trees and obtain the treewidth, 
a measure of the ''tree-likeness'' of the graph.  

Using dynamic programming techniques on tree decompositions of
bounded treewidth, hard optimization graph problems can often be
solved in polynomial time \cite{Arn91}.

The notions of treewidth and tree decomposition were introduced by
{\sc Robertson \& Seymour} in their seminal paper \cite{RobSey} on
graph minors.
The best known complexity bounds are given by the treewidth $tw$
({\sc Robertson, Seymour} \cite{RobSey}) of an interaction graph
associated with a DOP. This parameter is related to some topological
properties of the interaction graph. It leads to a time complexity of $O(n \cdot
2^{tw+1})$. Tree decomposition based methods aim to merge
variables so that the resulting meta-graph is a tree of meta-nodes. A more
detailed introduction to tree decompositions is given
in \cite{Kloks} and in surveys \cite{Bodl98}, \cite{Hicks}, \cite{Soa07t}. Most of the 
works based on tree decomposition approach only present theoretical
results \cite{Jegou_dyn}.
\begin{definition} Let $G=(V,E)$ be a graph. A \textbf{tree decomposition} of
$G$ is a pair $(T;X)$ with $T = (I; F)$ a tree and $X = \left\{
X_{i} \mid I \in I \right\}$ a family of subsets of $V$ , one for
each node of $T$, such that
\begin{itemize}
\item [(i)] $\bigcup_{i=1 \in I} X_i=V,$
\item [(ii)] for every edge $(u,v)\in V$ there is an $i\in I$ with $u \in X_i,~ v
\in X_i$,
\item [(iii)] (running intersection property) for all $i,j,l \in I$, if $i<j<l$, then
$X_i\cap X_l\subseteq X_j$.
\end{itemize}
\end{definition}
\begin{remark} To construct a tree-decomposition we merge the
vertices of $G$ together to form  meta-nodes (''supernodes''
\cite{Amestoy96} or ''bags'' \cite{Wolle}); each meta-node is a
subset of the vertices of $G$ and
 we connect these meta-nodes to form a rooted tree $T$. The meta-nodes do not have to be
disjoint, i.e., there might be nodes of the graph contained in
more than one meta-node. This grouping and connecting has to be done
in such a way that -- for all edges $e$ of the graph: there exists a
meta-node containing both endpoints of edge $e$, and -- for all
vertices $v$ of the graph: all meta-nodes containing $v$ together
with the edges between those meta-nodes in $T$ form a connected
subtree of $T$. 

The \textbf{width} of a tree-decomposition is the
number of graph-vertices of the largest meta-node minus 1.
\end{remark}
As finding an optimal tree-decomposition is $NP$-hard,
approximate optimal tree decompositions using triangulation of a
given graph are often exploited. Given a triangulated (or chordal)
graph, the set of its maximal cliques corresponds to the family of
subsets associated with a tree-decomposition (so called
\textbf{clique tree} \cite{Blair}). When
a tree-decomposition is exploited,  usually one considers approximations of optimal
triangulations by clique trees \cite{Jegou_comp}. Hence, the time complexity is then
$O(n \cdot 2^{w^{+}+1})$ with $w^{+}+1$
the size of the largest cluster ($w+1 \leq w^{+}+1 \leq n$). The space
complexity is $O(n \cdot s \cdot 2^{s})$ with $s$ the size of the
largest minimal separator \cite{Jegou_dyn}.

\begin{definition}
A \textbf{clique tree} of $G$ is a
tree $T = (K, E)$ whose vertex set is the set of maximal cliques of
$G$ such that each of the induced subgraphs $T[K_v]$ is connected.
\end{definition}
\begin{definition} A graph is \textbf{chordal}
(triangulated, perfect elimination graph, rigid circuit
\cite{dirac61}, monotone transitive \cite{Rose72}) if every cycle of
length $> 3$ has a chord (i.e., an edge joining
two nonconsecutive vertices of a cycle).
\end{definition}
All induced subgraphs of a
chordal graph are also chordal.
If $G$ is a chordal graph, then any clique tree of $G$ is also a
tree decomposition of $G$. However, the converse is not necessarily
true.
\begin{theorem}\label{thm:gavril} \cite{Gavril74}
 Let $G = (V,E)$ be an undirected graph, and
let $K$ be the set of maximal cliques of $G$, with $K_v$ the set of
all maximal cliques that contain vertex $v$ of $G$. The following
statements are equivalent:
\begin{itemize}
  \item [(i)] $G$ is chordal.
  \item [(ii)] $G$ is the intersection graph of a family
of subtrees of a tree.
  \item [(iii)] There exists a tree $T = (K, E)$ whose
vertex set is the set of maximal cliques of $G$ such that each of
the induced subgraphs $T[K_v]$ is connected.
\end{itemize}
\end{theorem}
Clique graphs are not very useful on general graphs, since these can
contain $n!/(k!(n - k)!)$ different cliques of size $k$. It follows
from this that clique graphs of general graphs can be exponentially
large, and it is no surprise that finding the maximal clique for
general graphs is hard. Chordal graphs on the other hand have
limitations that make clique graphs useful.
\begin{lemma} \label{dirac:2}({\sc Dirac} \cite{dirac61}) A chordal
graph $G$ contains at most $n$ maximal cliques.
\end{lemma}
%The clique graph may contain $O(n^2)$ edges.
{\sc Gavril}
\cite{Gavril74} proved that every chordal graph can be represented
by a clique tree limiting the number of edges to $O(n)$. Indeed,
from theorem \ref{thm:gavril} it follows that for the chordal
graph $G$ there exists a tree $T = (K, E)$ whose vertex set is the
set of maximal cliques of $G$ such that each of the induced
subgraphs $T[K_v]$ is connected. From Lemma \ref{dirac:2} and the
fact that $T$ is a tree, follows that a clique tree has at most $n$
nodes and $n - 1$ edges.
\begin{theorem} \label{bergood:1}({\sc
Bernstein and Goodman} \cite{BerGood81}) Any maximum weight spanning
tree of the clique graph of a chordal graph $G$ is a clique tree of
$G$.
\end{theorem}
\begin{theorem} \label{holee:1}({\sc Ho and Lee}
\cite{HoLee89}) Given a chordal graph $G$ and a clique tree $T$ of
$G$, a set of vertices $S$ is a minimal separator of $G$ iff $S =
C_i \cap C_j$ for an edge $(C_i,C_j)$ in $T$.
\end{theorem}
\begin{corollary} \label{holee:2}({\sc Ho and Lee} \cite{HoLee89}) A
chordal graph $G$ has at most $n - 1$ minimal separators.
\end{corollary}
These results show how to build tree decompositions using
elimination game algorithm which triangulates an initial interaction
graph. For triangulated graphs it is rather simply to find maximal
cliques and to build the clique tree.

\subsection{Elimination game and tree decomposition}
\label{sec_elim:6}

The process of interaction graph transformation known as {\it Elimination Game}
was first introduced by {\sc{Parter}} \cite{Part} as a graph analogy of
Gaussian elimination.  The input of the elimination game is a graph
$G$ and an ordering $\alpha$ of $G$ (i.e. $\alpha(v)=i$ if $v$ is
$i$-th vertex in ordering $\alpha$). Elimination Game according to
\cite{HEKP01} consists in the following.
 At each step $i$, the neighborhood of vertex $x_i$ is turned into a clique, and $x_i$ is deleted from the
graph. This is referred to as eliminating vertex $x_i$. 
The filled graph $G_{\alpha}^{+}=(V,E^+)$ is obtained by adding to
$G$ all the edges added by the algorithm.  This resulting  graph
$G_{\alpha}^{+}$ is a triangulation of $G$ ({\sc{FULKERSON \&
GROSS}} \cite{FulkGross}), i.e., a chordal graph.

Different filled graphs result from processing the vertices of $G$
in different orders. Thus in order to find a low fill-in, it is
necessary to find a good order on the vertices of the given graph
before running elimination game. Finding an ordering that results in
the minimum fill-in is an $NP$-complete problem \cite{Yanna}.\\
Elimination Game can also be implemented so that $\alpha$ is
generated during the course of the algorithm. In this case, we can
at each step $i$ choose a vertex $v$ of the elimination graph $G^{i-1}$ according to any
desired criteria, and set $\alpha(v)=i$, to define an elimination
ordering $\alpha$. One well known heuristic called Minimum Degree
chooses a vertex $v$ of minimum degree in $G^{i-1}$ at each step
$i$.
\begin{figure}[h!] \centering \includegraphics [height=10 cm]
{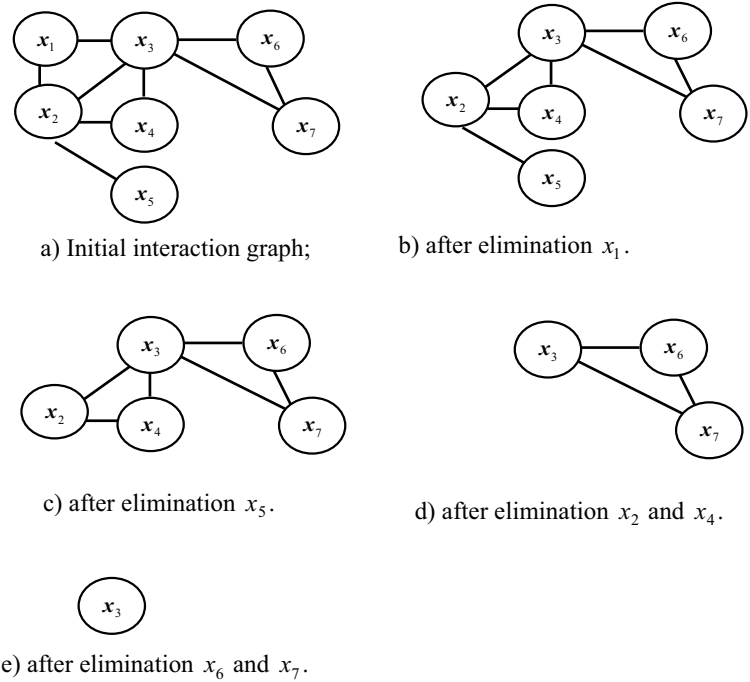} \caption{Elimination Game.} \label{ElimGame}
\end{figure}

The procedure to solve an optimization problem with bounded
treewidth involves two steps:
\begin{itemize}
  \item [(i)] computation of a (good) tree decomposition, and
  \item [(ii)]application of an (dynamic programming) algorithm that solves instances of
bounded treewidth in polynomial time.
\end{itemize}
To describe how tree-decompositions are used to solve problems with
dynamic programming, let us assume we find a tree-decomposition of a
graph $G$ \cite{Soa07k}. Since this tree-decomposition is represented as a rooted
tree $T$, the ancestor/ descendant relation is well-defined. We can
associate to each meta-node $X$ the subgraph of $G$ made up by the
vertices in $X$ and all its descendant meta-nodes, and all the edges
between those vertices. Starting at the leaves of the tree $T$, we
can compute information typically stored in a table, in a bottom-up
manner for each bag until we reach the root. This information is
sufficient to solve the subproblem for the corresponding subgraph.
To compute the table for a meta-node of the tree-decomposition, we
only need the information stored in the tables of the children (i.e.
direct descendants) of this node. The
problem for the entire graph can then be solved with the information
stored in the table of the root of $T$.

\section{Local decomposition algorithm for discrete optimization}
\label{s.la_dp_tree}

During the study of complex objects it is not always possible (and
expedient) to obtain (or to calculate) complete information about
the object as a whole; therefore is of interest to obtain
information about the object, examining it in parts, i.e., locally.

{\sc{Yu.I. Zhuravlev}}   \cite{ZhuR} introduced and investigated
local algorithms for calculating the information about properties of
objects. The local algorithm can be described as follows. For a
given set of sets $\left\{ m \right\},~m = \left\{ U_i
\right\},~i=1,2,\ldots,|m| $, and for each element $U \in m $ let us
determine a neighborhood $S(U,m)$ in $m$; these neighborhoods should
satisfy the following conditions:
 \begin{itemize}
\item $U \in S(U,m);$ \item $S(U,m) \sqsubseteq m;$ \item If $U \in
m_1, \, U \in m_2,\,S(U,m)  \sqsubseteq m_2  \sqsubseteq m_1$ then
$S(U,m_1)=S(U,m_2). $
 \end{itemize}

Algorithm $A$ is completely determined by the set of predicates
$P_1,\ldots,P_l$, by the partition of this set into a subset of
basic  predicates $P_1,\ldots,P_r$ and auxiliary  predicates
$P_{r+1},\ldots,P_l$ , by the set of monotonic functions
$\varphi_1,\ldots,\varphi_l, \: \varphi_i = \varphi_i(U,
\alpha_1,\ldots,\alpha_l,S,m^*)$ and by the ordering algorithm
$A_\pi$.

LA uses a dynamic programming paradigm and computes optimal partial
solutions of subproblems that correspond to the blocks of the DOP.

Each step of the LA $\mathfrak {A}$ \cite{Shch83} consists of
changing neighborhoods and replacing index $p$ with $p+1$
(although it is possible to pass, also, from $S_p$ to $S_{p+\rho}$);
for each fixed assignment of the variables of boundary ring the
values of the variables of the corresponding neighborhood are
stored, in this consists one of the important differences between LA
$\mathfrak {A}$ and LA $A$: an information not only about the
predicates, but also about the values of variables is memorized; this
information is called as an \textbf{indicator information}.

Let us consider the LA $\mathfrak{A}_{BT}$ for the solution BT  ILP
problems  (\ref{goal2}), (\ref{cons1}), (\ref{cons2}), where the matrix $A$ has
BT structure with the tree $D$ which contains of $k$ blocks.

Consider a vertex $r$ of the tree $D$ and introduce a tree $D_r$
which consists of the vertex $r$ and all its descendants.

Introduce the necessary notations:
\begin{itemize}
  \item $S_r$ is a  set of indices of variables which belong to block
$B_r$;
  \item $S_{rr^{'}}$ is a set of indices of variables which
belong simultaneously to blocks $B_r$ and $B_{r^{'}}$;
  \item if $S=\left\{ j_1,\,\ldots , j_q\right\}$, then $X_S=\left\{
x_{j_1},\ldots, x_{j_q}\right\}$;
  \item $p_r$ is a vertex-ancestor for the node $r$;
  \item $J_r$ is a set of descendants of the node $r$.
\end{itemize}
Consider a LA \cite{Shch83} for solving  DO problems with  a
tree structural graph, i.e., problems in which it is possible to find
the set of the neighborhoods of different variables so that one
variable can belong to two neighborhoods only and the graph of
intersections of these neighborhoods is a tree. The LA  solves this
DO problem, moving bottom-up, i.e., from the neighborhoods
corresponding to leaves of the tree, to the neighborhood
corresponding to the root of the tree $T$. Let $B_1=\left(
\bar{S}_1,~U_1 \right),~~ B_2=\left( \bar{S}_2,~U_2 \right),\ldots,
~~ B_k=\left( \bar{S}_k,~U_k \right)$ be a set of the blocks
(neighborhoods) of some indices $j_{1},\ldots,j_{k}$ of some
variables, where $ S_r,~U_r$ are, respectively, the sets of the
indices of variables and constraints for the $r$th neighborhood,
$r=1,\ldots,k$ and

\begin{equation} \label{u1}
 \bigcup _{r=1}^k U_r = M = \left\lbrace 1,\ldots, m\right\rbrace,\\
 \end{equation}
 \begin{equation} \label{s1}
\bigcup _{r=1}^k S_r = N = \left\lbrace 1,\ldots, n\right\rbrace,
\end{equation}
\begin{equation} \label{u1u2} U_{r_1}\cap U_{r_2} =
\emptyset,~ r_1\neq r_2,
\end{equation}
\begin{equation}
\label{s1s2s3} S_{r_1}\cap S_{r_2}\cap S_{r_3} = \emptyset \mbox {
for any triple of different indices }  r_1, r_2, r_3 .
\end{equation}
Consider a node $r$ of the tree $T$ and define a tree $T_r$ containing
the node $r$ and its children.

Introduce notions:
\begin{itemize}
  \item $\bar{S}_r$ is a set of indices of variables that are in the block
$B_r$;
  \item $S_{rr^{'}}$ is a set of indices of variables that are in the blocks
   $B_r$ and $B_{r^{'}}$,
i.e., $S_{rr^{'}}=\bar{S}_r \bigcap \bar{S}_{r^{'}}$;
  \item $p_r$ is a node-parent of the node
â $r$;
 \item $J_r$ is a set of children of the node $r$.
\end{itemize}
Then $X_{S_{{p_r}r}}$ is a meta-variable consisting of variables common
for blocks $B_{p_r}$ and $B_r$ (here $S_{{p_r}r}=\bar{S}_{p_r} \bigcap
\bar{S}_{r}$).

Denote as $Z_{T_r}$ the following problem: for each assignment
 $X_{S_{{p_r}r}}$ to find $X_{S_r}$ è $X_{S_{rr^{'}}}$, such that\\
 \[
 h_{B_r}(X_{S_{{p_r}r}}) = f_{D_{r}}\left(X_{S_{p_{r}r}}\right)= \max_{X_{S_r},X_{S_{rr^{'}}}} \left\{ C_{S_{r}}X_{S_{r}}+
 \sum_{r^{'} \in J_{r}} \left[f_{D_{r^{'}}}(X_{S_{rr^{'}}}) + C_{S_{rr^{'}}}X_{S_{rr^{'}}} \right] \right\} =
 \]
\[= \max_{X_{S_r},X_{S_{rr^{'}}}} \left\{ C_{S_{r}}X_{S_{r}}+
 \sum_{r^{'} \in J_{r}} \left[h_{B_{r^{'}}}(X_{\bar{S}_r \bigcap \bar{S}_{r^{'}}}) + C_{S_{rr^{'}}}X_{S_{rr^{'}}} \right] \right\}
\] subject to \[
 A_{S_{r}}X_{S_{r}} \leq b_{r} - \sum_{r^{'} \in J_{r}} A_{S_{rr^{'}}}X_{S_{rr^{'}}} -
 A_{S_{p_{r}r}}X_{S_{p_{r}r}} .
\]

Here $f_{T_{r}}(X_{S_{p_{r}r}})$ is an objective value of subproblem
corresponding to the tree $T_{r}$,
$f_{T_{r^{'}}}\left(X_{S_{rr^{'}}}\right)=C_{T_{r^{'}}}X_{T_{r^{'}}}$.
It is possible to assign this value to the root $B_{r^{'}}$ of the tree $T_{r}$ and
write: $h_{B_{r^{'}}}(X_{\bar{S}_r \bigcap
\bar{S}_{r^{'}}})=f_{T_{r^{'}}}\left(X_{S_{rr^{'}}}\right)$.

It is easy to see that if we fix a vector $X_{S_{p_{r}r}}$, then the
problem is decomposed into two problems: the first one corresponds
to the tree $T_{r}$; and the second one to $T - T_{r}$. An
application of LA $\mathfrak{A}_{BT}$ for solving DP problems with a
tree structural graph is based on this property.

\bigskip \section{Solving of a concrete DOP with the finding of tree
like structure and applying of local decomposition algorithm}
\label{s.la_primer}

Consider the DOP of example 1.

 \textbf{Finding of the tree structural graph (tree decomposition)}

\begin{figure}[ht]
\centering \includegraphics [height=14 cm]
{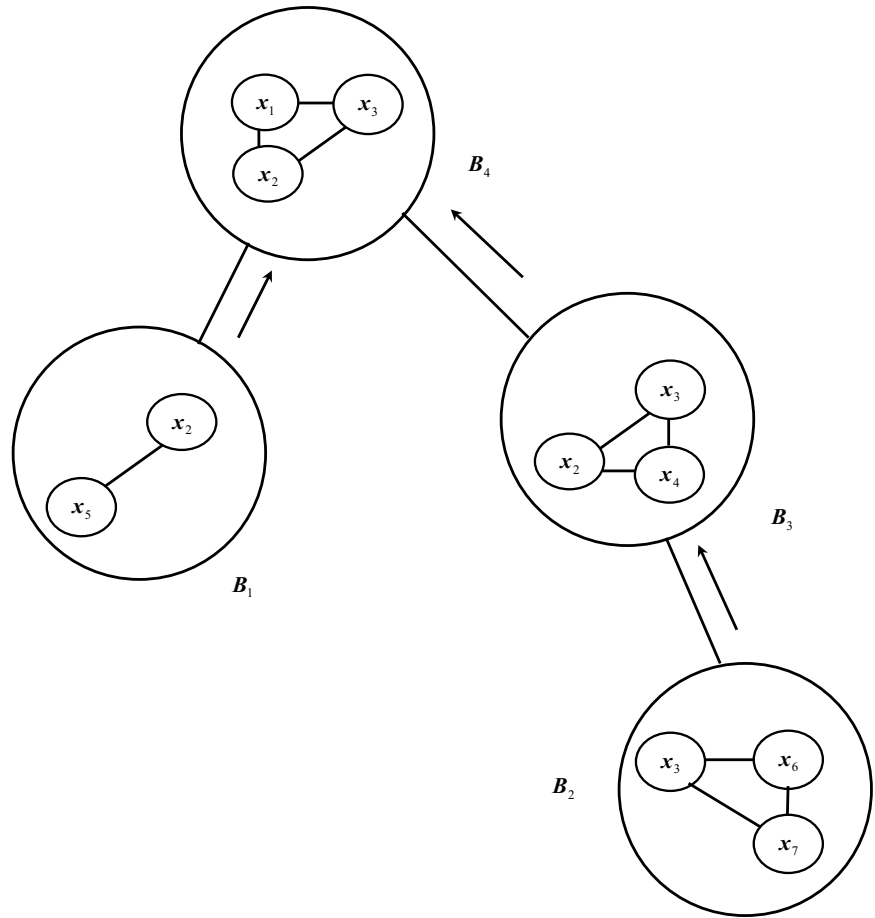}
\caption{Tree decomposition for the example 1.}
\label{dd_pr}
\end{figure}

In Fig. \ref{ElimGame} results of elimination game algorithm are
shown. Since during elimination process new fill-in edges are not
added, the elimination game process is equivalent to searching
simplicial vertices and corresponding maximal cliques. In Fig.
\ref{dd_pr} these maximal cliques and links between them are shown.
Local decomposition algorithm can be applied to this clique tree.
Other possible way of finding of the clique tree is using of maximal
spanning tree in the dual graph.

\textbf{Applying the local decomposition algorithm to DO problem}

Let us solve the subproblem corresponding to the block $B_1$. Since this
block is adjacent to the block $B_4$, we have to solve DOP with
variables $X_{B_1-B_4}$ for all possible assignments $X_{B_1 \bigcap
B_4}$. Thus, since $X_{B_1-B_4}=\left\{ x_5\right\}$ and $X_{B_1
\bigcap B_4}=\left\{ x_2\right\}$, then induced subproblem has a
form: \[ h_{B_1}(x_2)=\max_{x_5}\left\{4x_5\right\} \] subject to \[
2x_2+3x_5 \le 4,~~x_j=0,1,~j \in \left\{2,5\right\} \] \bigskip

Solution of the problem can be written in tabular form:\\

%\begin{table}[ht] %\caption{

\begin{center} Table 1. Calculation of $h_{B_1}(x_2)$\\ \smallskip
%} \label{t.table1}
\begin{tabular}{|c|c|c|} \hline
 $x_2$&   $h_{B_1}$ & $x_5^*(x_2)$ \\
\hline 0 &   4 &   1 \\ 1 &   0 &   0 \\ \hline \end{tabular}
\end{center} %\end{table} \bigskip

Next subproblem corresponding to a leaf (or meta-node) $B_2$ of the
clique tree is

\[ h_{B_2}(x_3)=\max_{x_6,x_7}\left\{6x_6+x_7\right\} \] s.t. \[
2x_3+3x_6+2x_7 \le 5,~~x_j=0,1,~j \in \left\{3,6,7\right\} \]
Solution of this subproblem:

\bigskip %\begin{table}[htbp] %\caption{

\begin{center} Table 2. Calculation of $h_{B_2}(x_3)$\\%}
\label{t.table2} \smallskip \begin{tabular}{|c|c|cc|} \hline
 $x_3$ & $h_{B_2}$ & $x_6^*(x_3)$ & $x_7^*(x_3)$ \\
\hline 1 &1&   7 &   1 \\ 1 &0&   6 &   1 \\ \hline \end{tabular}
\end{center} \bigskip %\end{table}\\

Subproblem corresponding to the block $B_3$ has the form: 
\[
h_{B_3}(x_2,x_3)=\max_{x_4}\left\{h_{B_2}(x_3)+5x_4\right\} \] s.t.
\[ 2x_2+3x_3+3x_4 \le 5,~~x_j=0,1,~j \in \left\{2,3,4\right\} \]

%\begin{table}[htbp] %\caption{
\begin{center} Table 3. Calculation
of $h_{B_3}(x_2,x_3)$ \\%} \label{t.table3} \smallskip
\begin{tabular}{|cc|c|c|} \hline
 $x_2$ & $x_3$ & $ h_{B_3}$ & $x_4^*(x_2,x_3)$ \\
\hline 0 &0&   12 &   1 \\ 0 &1&   6 &   0 \\ 1 &0&   12 &   1 \\ 1
&1&   6 &   0 \\ \hline \end{tabular} \end{center} %\end{table}\\
\bigskip

The last problem left to be solved is: \[
h_{B_4}=\max_{x_1,x_2,x_3}\left\{h_{B_1}(x_2)+ h_{B_3}(x_2,x_3)+
2x_1\right\} \] s.t.
 \[ 3x_1+4x_2+x_3 \le 6,~~x_j=0,1,~j \in
\left\{1,2,3\right\} \] %\begin{table}[htbp] %\caption{

\begin{center} Table 4. Calculation of $h_{B_4}$ \\%}
\label{t.table4} \smallskip \begin{tabular}{|c|ccc|} \hline
 $ h_{B_4}$ & $x_1^*$ & $x_2^*$ & $x_3^*$ \\
\hline 18 &1&   0 &   0 \\ \hline \end{tabular} \end{center}
\bigskip %\end{table}\\

The maximal objective value is 18. To find the optimal values of the
variables, it is necessary to do backward step of the dynamic
programming procedure: from  table 4 we have $x_1^*=1,~ x_2^*=0,
~x_3^*=0$. From  table 3 using the information $ x_2^*=0, ~x_3^*=0$
we find $ x_4^*=1$. Considering  table 2  we have for $x_3^*$=0:
$x_6^*=1,x_7^*=1$. From  table 1 we find for $ x_2^*=0$: $
x_5^*=1$. The solution is (1, 0, 0, 1, 1, 1, 1); maximal objective
value is 18.

\section{Postoptimal analysis and local algorithms}

\subsection{Postoptimal analysis in DO}

Decomposition and sensitivity analysis in DO are closely related. Sensitivity analysis follows
naturally from the duality theory. Decomposition methods consist of generating and solving
families of related DO problems that have the same structure but differ as the values of coefficients.
Sensitivity analysis allows using information obtained during solving one DO problem of the family
of related DO problems in solving other problems of this family. Due to the lack of full-fledged
duality theory in DO, sensitivity analysis for DO problems is not sufficiently developed \cite{GeNa77},
 \cite{MaMo77}. A number of useful tools of sensitivity analysis in DO are derived for
integer programming in \cite{GeNa77}. A technique of sensitivity analysis proposed in \cite{SchWo85}
computes a piecewise linear value function that provides a lower bound on the optimal value that
results from changing the right-hand sides of constraints.

Recently, an interesting application of binary decision diagrams (BDD)
(introduced earlier in computer science community) was proposed by
{\sc Hadzic \& Hooker}  \cite{HadHo07} for the
purposes of postoptimal analysis in DO.

\subsection{Postoptimal analysis in LA}
LA systematically proceeds with so called \textbf{parametric} DO
problems \cite{BerBri}:
\begin{definition}
A parametric DO problem is
\[
\min_{X-P}\{f(X)|X \in C_i,~i=1,\ldots,m;~x_j \in \{0,1\},~j=1,\ldots,n\},
\]
where $C_i$ is a set of feasible solutions of the constraint $i, ~i=1,\ldots,m$.
\end{definition}
Thus, an optimization problem is in \emph{parametric
form} when the objective function is optimized not over the entire set
$X$, but only over a subset $X-P$, for all possible assignments of
 the variables of $P$.

Below we show that these parametric DOPs generated by LA lead to the possibility of
exploiting postoptimality and sensitivity tools in the LA procedure.
Consider  DOP (\ref{goal2}), (\ref{cons1}), (\ref{cons2}) with tree
structural graph and use LA for its solving. Then for
the block $B_r$ we have to solve a family of DOPs:
\[
C_{S_r}X_{S_r}+[C_{S_{r-1,r}}X_{S_{r-1,r}}+C_{\sigma_{r-1}}X_{\sigma_{r-1}}]
\rightarrow \max \]
s.t.
\[
\sum_{j \in S_r} a_{ij}x_j \leq b_i -
A^{i}_{S_{r-1,r}}X_{S_{r-1,r}}-A^{i}_{S_{r,r+1}}X_{S_{r,r+1}},~i \in
U_r.
\]
These DOPs should be solved for all binary assignments $X_{S_{r-1,r}}$ and $X_{S_{r,r+1}}$.
Denote
\[
b_i \left(X_{S_{r-1,r}}| X_{S_{r,r+1}} \right) = b_i -
A^{i}_{S_{r-1,r}}X_{S_{r-1,r}}-A^{i}_{S_{r,r+1}}X_{S_{r,r+1}}.
\]
It is clear, that it is better for each $X_{S_{r-1,r}}$ and $X_{S_{r,r+1}}$ to solve a problem:

\begin{equation}\label{obj}
C_{S_r}X_{S_r} \rightarrow \max
\end{equation}
s.t.
\begin{equation}\label{Ax}
\sum_{j \in S_r} a_{ij}x_j \leq b_i \left(X_{S_{r-1,r}}|
X_{S_{r,r+1}} \right),~i \in U_r,
\end{equation}
\begin{equation}\label{01}
x_j = 0,1,~ j \in S_r.
\end{equation}
It is possible to use information obtained during the solution of some DO
problems of the family (\ref{obj}--\ref{01}) for solving  other problems of this family using
postoptimality analysis (PA) \cite{GeNa77}, \cite{Green97},
 \cite{HadHo07}, \cite{SchWo85}. The more efficient
the PA procedure is, the better LA will work.

\subsection{Postoptimal analysis for an implicit enumeration algorithm}

We show how to use PA in LA using as an illustrative example, the simplest
case of an implicit enumeration algorithm \cite{Geoff} that generates partial solutions
and tries to fathom them using 3 tests.

The DO problem (\ref{obj})--(\ref{01}) with vector right hand side $b_i
\left(X_{S_{r-1,r}}| X_{S_{r,r+1}} \right)$ will be denoted as $Z_r
\left(X_{S_{r-1,r}}| X_{S_{r,r+1}} \right)$. Introduce a partial
order over a set of DO problems $\{Z_r \left(X_{S_{r-1,r}}|
X_{S_{r,r+1}} \right)\}$: DOP $Z_{r}^{'}=Z_r
\left(X_{S_{r-1,r}}^{'}| X_{S_{r,r+1}}^{'} \right)$ precedes DOP
$Z_{r}^{''}=Z_r \left(X_{S_{r-1,r}}^{''}| X_{S_{r,r+1}}^{''}
\right)$ if $b_i \left(X_{S_{r-1,r}}^{'}| X_{S_{r,r+1}}^{'} \right)
\leq b_i \left(X_{S_{r-1,r}}^{''}| X_{S_{r,r+1}}^{''} \right), ~i
\in U_r$ or graphically

\begin{figure}[htbp] \centering
\includegraphics[height=5cm]{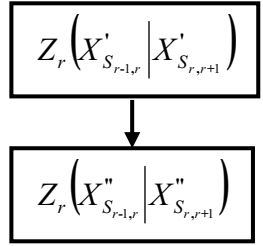} \caption{Partial order of DOPs.} \label{ord}
\end{figure}

Let us solve the problem $Z_{r}^{'}$ by  implicit enumeration.
Let P be a partial solution (PS) fathomed by the tests of the
implicit enumeration. There are 3 possible cases:
\begin{itemize}
  \item [a)] PS $P$ is fathomed by test 2, i.e., the best completion of $P$
  is feasible. Then this completion is feasible for the problem  $Z_{r}^{''}$
  too, so as
  \[
  b_i \left(X_{S_{r-1,r}}^{'}| X_{S_{r,r+1}}^{'} \right)
\leq b_i \left(X_{S_{r-1,r}}^{''}| X_{S_{r,r+1}}^{''} \right).
  \]
  Thus, each incumbent of $Z_{r}^{'}$ is a feasible solution of $Z_{r}^{''}$,i.e.,
 the objective function value of  $Z_{r}^{''}$ is higher (better) than the corresponding
  objective function value of  $Z_{r}^{'}$.
  \item [b)] PS $P$ is fathomed by  test 1 in the problem $Z_{r}^{'}$, i.e.,
  $\overline{z} \leq z^{'*}$. Then $P$ is fathomed in $Z_{r}^{''}$, too, as
  $\overline{z} \leq z^{'*} \leq z^{''*}$.
  \item [c)] PS $P$ is fathomed by test 3 in $Z_{r}^{'}$.
\end{itemize}

It is clear, that in cases a) and b) it is senseless to fathom PS
$P$ in the problem $Z_{r}^{''}$, because $P$ is automatically
fathomed. Thus, while fathoming PS in $Z_{r}^{''}$, it is
interesting to study only those PS $P$ fathomed by  test 3 in
$Z_{r}^{'}$.

\begin{figure}[htbp]
\centering
\includegraphics[height=8cm]{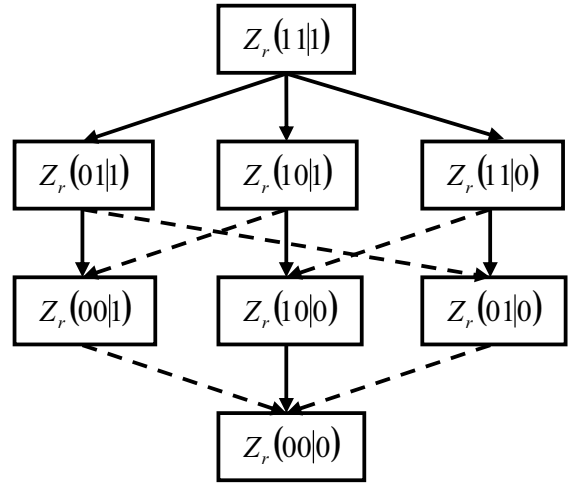}
\caption{Partial order of DOPs (multidimensional
knapsack).} \label{part_order}
\end{figure}

Let us consider a family of $2^{|S_{r-1,r}|+|S_{r,r+1}|}$ DOPs
(\ref{obj})--(\ref{01}) with nonnegative coefficients (multidimensional
knapsack problems). Using  partial order of DOPs described above
it is possible
to order members of this family as it shown in Fig.
\ref{part_order}. Consider a process of solving the family of DOPs
$\{Z_r\}$. Let $P$ be a PS fathomed in $\{Z_r(11|1)\}$ (level 0 in
Fig. \ref{part_order}). If PS $P$ is fathomed by tests 1, 2, we can
exclude this PS $P$ from consideration (and fathoming) in other
DOPs.
If PS $P$ was fathomed by the test 3 in $\{Z_r(11|1)\}$, then we
pass to a DOP from the level 1, say, $\{Z_r(01|1)\}$ and consider
the PS $P$ in this problem.

If the PS $P$ is not fathomed by the test 3 in $\{Z_r(01|1)\}$, then there
exists one of the following cases: 
\begin{itemize}
  \item test 1 is true, then backtrack to $\{Z_r(11|1)\}$ and pass to
  the next problem of  level 1;
  \item test 2 is true, then also backtrack to $\{Z_r(11|1)\}$ and pass to
  the next problem of  level 1;
  \item both tests 1 and 2 are not true, then extend PS $P:~P'=(P,j_1)$ and try
  to fathom $P'$.
\end{itemize}
If any of the tests 1, 2 are true, then do usual backtracking, i.e.,
PS $P''=(P,-j_1)$ is considered. If some extension of
$P:~P'''=(P,j_1,-j2,\ldots,j_f)$ is fathomed by the test 3, then we
go to one of problem's ancestors (say, $\{Z_r(00|1)\}$) and try to
fathom by the tests of implicit enumeration.

 \section*{Conclusion} Local decomposition algorithms combined
 with tree decomposition methods are a promising approach that enables
 solving  sparse discrete optimization problems from applications. The  performance
 of these algorithms can be improved with the aid of postoptimality analysis.

\textbf{A promising direction of future research} is the development for efficient
schemes of postoptimality analysis embedded in local decomposition algorithms combined
 with tree decomposition methods.


\medskip \begin{thebibliography}{3}

\bibitem{Amestoy96} Amestoy, P.R.,  Davis, T.A.,  Duff, I.S.: An
approximate minimum degree ordering algorithm. SIAM J. on
Matrix Analysis and Appl. {\bf 17}, 886--905 (1996)

\bibitem{Arn87}  Arnborg, S.,  Corneil, D.G., Proskurowski, A.:
Complexity of finding embeddings in a k-tree. SIAM J. of Alg. Discrete
Methods {\bf 8}, 277--284 (1987)

\bibitem{Arn91}  Arnborg, S.,  Lagergren, J.,  Seese, D.: Easy
problems for tree-decomposable graphs. J. of Algorithms {\bf 12}, 308--340 (1991)

\bibitem{BerGood81} Bernstein, P.A.,  Goodman, N.:  Power of natural
semijoins. SIAM J. on Computing  {\bf 10}, 751--771 (1981)

%\bibitem{BerBl}Berry, A.,  Blair, J.,  Heggernes, P., Peyton, B.:
%Maximum cardinality search for computing minimal triangulations of
%graphs. Algorithmica {\bf 39}, 287--298 (2004)

\bibitem{BerBri} Bertele,  U.,  Brioschi, F.: Nonserial Dynamic
Programming.  Academic Press, New York (1972)

\bibitem{Blair} Blair, J.R.S.,  Peyton, B.W.:  An introduction to
chordal graphs and clique trees. In:  George, J.A. et al. (eds.)
  Sparse Matrix Computations: Graph Theory Issues
and Algorithms. Springer Verlag, p.1--30 (1993)

\bibitem{Bodl93} Bodlaender, H.L.:  A tourist guide through
treewidth. Acta Cybernetica  {\bf 11}, 1--21 (1993)

\bibitem{Bodl98} Bodlaender, H.L.: A partial k-arboretum of graphs
with bounded treewidth. Theoretical Computer Science  {\bf 209},
1--45 (1998)

\bibitem{bodl} Bodlaender, H.L.:  Discovering Treewidth.  In:
Proceedings of SOFSEM, LNCS {\bf 3381}. Springer-Verlag, p.1--16 (2006)

%\bibitem{Bryant86}  Bryant RE.   Graph-based algorithms for boolean function manipulation. IEEE
%Transactions on Computers 1986; C-35, 8; 677-691.

\bibitem{cook94} Cook, W.,  Seymour, P.D.:  An algorithm for the
ring-routing problem. Bellcore technical memorandum, Bellcore (1994)

\bibitem{cook03} Cook, W.,  Seymour, P.D.:  Tour merging via
branch-decomposition. INFORMS J. on Computing {\bf 15}, 233--248 (2003)

\bibitem{cour}  Courcelle, B.: Graph rewriting: An algebraic and
logic approach. In:  Van Leeuwen, J. (ed.) Handbook of Theoretical
Computer Science, Volume B. Elsevier,  p.193--242 (1990)

\bibitem{Decht92} Dechter, R.: Constraint networks. In: Encyclopedia
of Artificial Intelligence, 2nd edn. Wiley, New York, p.276--285 (1992)

%\bibitem{Dechter} Dechter, R.: Bucket elimination: A unifying
%framework for reasoning. Articial  Intelligence {\bf 113}, 41--85
%(1999)
%
\bibitem{Dechter2001}  Dechter, R.,   El Fattah, Y.: Topological
parameters for time-space tradeoff. Artif. Intell.  {\bf 125},
93--118 (2001)

\bibitem{dirac61}  Dirac, G.A.:  On rigid circuit graphs. Abhandlungen Mathematischen
Seminar Universit\"{a}t Hamburg  {\bf 25}, 71--76 (1961)

%\bibitem{Fink65}
% Finkel'shtein, Yu.Yu.: On solving discrete programming problems of
%special form (Russian). Economics and Math. Methods {\bf 1},
%262--270 (1965)
%
\bibitem{FulkGross}  Fulkerson, D.R., Gross, O.A.:  Incidence
matrices and interval graphs. Pacific J. of Mathematics {\bf 15},
835--855 (1965)

\bibitem{Gavril74} Gavril, F.:  The intersection graphs of subtrees
in trees are exactly the chordal graphs. J. of Comb. Theory
Ser.B. {\bf 16}, 47--56 (1974)

\bibitem{Geoff} Geoffrion, A.M.:  An improved implicit
enumeration approach for integer programming. Operations Research {\bf 17}, 437--454 (1969)


\bibitem{GeNa77}  Geoffrion, A.M.,  Nauss, R.:  Parametric and postoptimality
analysis in integer linear programming. Management Science {\bf 23},
453--466 (1977)

%\bibitem{george} George, J.A.,  Liu, J.W.H.:  Computer Solution of
%Large Sparse Positive Definite Systems. Prentice-Hall Inc.,
%Englewood Cliffs (1981)
%
%\bibitem{Gott01}  Gottlob G,   Leone N, Scarcello F. Hypertree
%decompositions: A survey. In: Mathematical foundations of computer
%science. 26th international symposium, MFCS 2001, Proceedings, LNCS 2136.
% Springer: Berlin; 2001. p.37-57.


\bibitem{Green97}  Greenberg, H.J.:   An annotated bibliography for post-solution
analysis in mixed integer programming and combinatorial
optimization. In: Advances in computational and stochastic optimization,
logic programming, and heuristic search: interfaces in computer
science and operations research. Kluwer Academic Publishers,
Norwell, MA (1997)

%\bibitem{HadHo06}  Hadzic T,  Hooker J.  Postoptimality analysis for integer
%programming using binary decision diagrams. Carnegie Mellon
%University (Presented at GICOLAG workshop, Vienna); 2006.

\bibitem{HadHo07}  Hadzic, T.,  Hooker, J.:  Cost-bounded binary decision diagrams
for 0-1 programming. In:  Loute, E.,  Wolsey, L. (eds.), Integration of
AI and OR techniques in Constraint Programming for combinatorial
optimization problems (CPAIOR 2007) (2007)


%\bibitem{hajnal}  Hajnal, A., Suranyi, J.:  \"{U}ber die
%Ausfl\"{o}sung von Graphen in vollst\"{a}ndige Teilgraphen. Ann.
%Univ. Sci. Budapest. 113--121 (1958)
%
%\bibitem{Harary} Harary, F., Norman, R.Z., Cartwright, D.:
%Structural Models: An Introduction to the Theory of Directed Graphs.
%John Wiley \& Sons (1965)

\bibitem{Hegg} Heggernes, P.: Treewidth, partial k-trees, and
chordal graphs. \\ URL: {\tt http://www.ii.uib.no/~pinar/chordal.pdf}

%\bibitem{Heggm}  Heggernes, P.: Minimal triangulations of graphs: A
%survey. Discrete Mathematics {\bf 306}, 3, 297--317 (2006)
%
\bibitem{HEKP01} Heggernes, P., Eisenstat, S.C., Kumfert, G.,
Pothen, A.: The Computational Complexity of the Minimum Degree
Algorithm. Techn. report UCRL-ID-148375. Lawrence Livermore National
Laboratory (2001) \\ URL:
{\tt http://www.llnl.gov/tid/lof/documents/pdf/241278.pdf}


\bibitem{Hicks}  Hicks, I.V.,  Koster, A.M.C.A.,  Kolotoglu, E.:
Branch and tree decomposition techniques for discrete optimization.
 In: Tutorials in Operations Research.  INFORMS, New Orleans (2005)\\
 URL: {\tt http://ie.tamu.edu/People/faculty/Hicks/bwtw.pdf}

\bibitem{HoLee89} Ho, C.W.,  Lee, C.T.:  Counting clique trees and
computing perfect elimination schemes in parallel. Information Processing
Letters {\bf 31}, 61--68 (1989)

%\bibitem{Hook} Hooker J. Logic-based methods for optimization:
%combining optimization and constraint satisfaction.  John Wiley;
% 2000. %. -- 495 p.

\bibitem{Jegou_comp} J\'{e}gou, P.,  Ndiaye, S.N., Terrioux C.:
Computing and exploiting tree-decompositions for (Max-)CSP. In:
Proceedings of the 11th International Conference on Principles and
Practice of Constraint Programming (CP-2005), p.777--781 (2005)

\bibitem{Jegou_dyn} J\'{e}gou, P.,  Ndiaye, S.N., Terrioux C.:
Dynamic heuristics for backtrack search on tree-decompositions of
CSPs. In: Proceedings of the 20th International Joint Conference on
Artificial Intelligence (IJCAI-07) (2007)

\bibitem{Kloks} Kloks, T.:  Treewidth: Computations and
Approximations. Lecture Notes in Computer Science {\bf 842},
Springer-Verlag (1994)

\bibitem{Koster99} Koster, A.M.C.A.,  van
Hoesel, S.P.M., Kolen, A.W.J.: Solving frequency assignment problems via tree-decomposition. In:
Broersma, H.J. et al. (eds.) 6th Twente workshop on graphs and
combinatorial optimization. Univ. of Twente, Enschede, Netherlands,
May 26-28, 1999. Extended abstracts. Elsevier, Amsterdam. Electron.
Notes Discrete Math. {\bf 3}, no pag., electronic only (1999)

\bibitem{kost01}  Koster, A.M.C.A.,  Bodlaender, H.L.,  van
Hoesel, S.P.M.: Treewidth: Computational experiments. In:  Broersma H et al. (eds.)
Electronic Notes in Discrete
Mathematics, Elsevier Science Publishers {\bf 8} (2001)
%
%\bibitem{lewis} T.G. Lewis, B.W. Peyton, A. Pothen.  A fast
%algorithm for reordering %sparse matrices for parallel factorization
%// SIAM J. Sci. Stat. %Comput. -- 1989. -- V. 10. -- P. 1146-1173.
\bibitem{MaMo77}  Marsten, R.E.,  Morin T.L.:  Parametric integer programming: The right
hand side case. In:  Hammer, P. et al. (eds.) Annals of Discrete
Mathematics: Studies in Integer Programming. Elsevier, North Holland,
 p. 375--390 (1977)

\bibitem{Neus}  Neumaier, A.,  Shcherbina, O.: Nonserial dynamic
programming and local decomposition algorithms in discrete
programming (submitted). Available online:\\
{\tt http://www.optimization-online.org/DB\_HTML/2006/03/1351.html}

\bibitem{Part} Parter, S.: The use of linear graphs in Gauss
elimination. SIAM Review {\bf 3}, 119--130 (1961)

\bibitem{RalG05} Ralphs, T.K.,  Galati, M.V.:   Decomposition in
integer linear programming. In:  Karlof, J. (ed.) Integer
Programming: Theory and Practice  (2005) \\URL:
{ \tt http://www.optimization-online.org/DB\_HTML/2004/12/1029.html}

\bibitem{RobSey} Robertson, N.,  Seymour, P.D.: Graph minors. II.
Algorithmic aspects of tree width. J. of Algorithms {\bf 7}, 309--322 (1986)

\bibitem{Rose72} Rose, D.J.:  A graph-theoretic study of the
numerical solution of sparse positive definite systems of linear
equations. In: Read, R.C. (ed.) Graph Theory and Computing.
Academic Press, New York, p.183--217 (1972)

\bibitem{SchWo85} Schrage, L.E.,  Wolsey, L.A.:  Sensitivity analysis for branch
and bound integer programming. Operations Research {\bf 33}, 1008--1023 (1985)

\bibitem{Shch83}  Shcherbina, O.A.: On local algorithms of solving
discrete optimization problems. Problems of Cybernetics (Moscow) {\bf 40}, 171--200 (1983)
 .

\bibitem{Soa07k} Shcherbina, O.: Nonserial dynamic programming and
tree decomposition in discrete optimization. In: Proc. of Int.
Conference on Operations Research "Operations Research 2006".
Karlsruhe, 6-8 September, 2006, p.155-160 Springer Verlag, Berlin
(2007)

\bibitem{Soa07t}  Shcherbina, O.A.: Tree decomposition and discrete optimization problems:
 A survey. Cybernetics and Systems Analysis {\bf 43}, 549--562 (2007)  %URL:
% http://www.springerlink.com/content/n6w880rw24513433/?p=50c98ca390f3487fa0d01ff9927c67d6&pi=7


%\bibitem{Shib} Shibata, Y.:  On the tree representation of chordal
%graphs. J. Graph Theory {\bf 12}, 421--428 (1988)
%
\bibitem{TarYan}  Tarjan, R.E., Yannakakis, M.:  Simple linear-time
algorithms to test chordality of graphs, test acyclity of
hypergraphs, and selectively reduce acyclic hypergraphs. SIAM J.
on Computing {\bf 13}, 566--579 (1984)

%\bibitem{Tsang93} Tsang  E. Foundations of Constraint
%Satisfaction. Academic Press: New York; 1993. %-- 421 p.

\bibitem{Wolle} Wolle, T.:  Computational Aspects of Treewidth,
Lower Bounds and Networks Reliability. Dissertation, UU Universiteit
Utrecht (2005) \\URL:
{\tt http://igitur-archive.library.uu.nl/dissertations/2005-0614-200103/index.htm}

\bibitem{Yanna}  Yannakakis, M.: Computing the minimum fill-in is
NP-complete. SIAM J. of Alg. Discrete
Methods {\bf 2}, 77--79 (1981)

\bibitem{ZhuR} Zhuravlev, Y.I.: Local algorithm of information
computation (Russian), I,II. Kibernetika {\bf 1}, 12--19 (1965); {\bf 2}, 1--11 (1966)

%\end{thebibliography} \end{document}

%\bibitem{Ref1}
%Author, I.: Article title. Journal Title-Abbreviated {\bf Vol}, pp--pp (year)
%\bibitem{Ref2}
%Author, I., Smith, J.: Book Title. Publisher, Place (year)
% Format for proceedings
%\bibitem{Ref3}
%Author, I., Smith, J.: Paper title. In: Editor, A. (ed.) Proceedings
%Title, Location, Date, pages. Publisher, Place (year)
% etc
\end{thebibliography}
\end{document}